# Deterministic non-local parity control and supercurrent-based detection in an Andreev molecule

Shang Zhu[1,2,#], Xiaozhou Yang[1,2,#], Mingli Liu[1], Min Wei[1,2], Yiping Jiao[1,2], Jiezhong He[1,2], Bingbing Tong[1,3], Junya Feng[1], Ziwei Dou[1], Peiling Li[1,3], Jie Shen[1], Xiaohui Song[1,3], Guangtong Liu[1,3], Zhaozheng Lyu[1,3,*], Dong Pan[4,*], Jianhua Zhao[5], Li Lu[1,2,3,*], Fanming Qu[1,2,3,*]

[1] Beijing National Laboratory for Condensed Matter Physics, Institute of Physics, Chinese Academy of Sciences, Beijing 100190, China
[2] University of Chinese Academy of Sciences, Beijing 100049, China
[3] Hefei National Laboratory, Hefei 230088, China
[4] State Key Laboratory of Semiconductor Physics and Chip Technologies, Institute of Semiconductors, Chinese Academy of Sciences, Beijing 100083, China
[5] State Key Laboratory of Spintronics, Hangzhou International Innovation Institute, Beihang University, Hangzhou 311115, China
[#] These authors contributed equally to this work.
[*] Email: lyuzhzh@iphy.ac.cn; pandong@semi.ac.cn; lilu@iphy.ac.cn; fanmingqu@iphy.ac.cn

**Abstract**

The ability to manipulate and detect the parity of quantum states in superconductor–semiconductor hybrid systems is pivotal to realizing the promise of topological quantum computation. However, as these architectures scale toward artificial Kitaev chains with phase-control loops, local accessibility becomes restricted, constraining conventional local parity control and detection. While Andreev molecules offer a platform for non-local intervention, deterministic protocols for parity manipulation have yet to be experimentally established. Here, we demonstrate deterministic non-local control over the parity configuration of a quantum dot (QD) by electrically modulating the coherent hybridization with a spatially adjacent QD within an Andreev molecule. By systematically investigating three distinct joint parity configuration regimes in the elastic co-tunneling limit, we experimentally uncover the operational conditions for this non-local control. In conjunction with theoretical simulations establishing a global phase diagram, we identify a set of universal selection rules governing parity transitions, dictated by the symmetry-imposed interplay between the joint parity configuration and the dominant inter-dot coupling mechanism (elastic co-tunneling vs. crossed Andreev reflection). Furthermore, we establish the supercurrent, directly signaled by zero-bias conductance peaks, as an intrinsic, sensor-free probe of the parity configuration, obviating the need for auxiliary charge sensors. Our results provide a validated physical framework for parity engineering, offering a key building block for scalable, multi-QD superconducting architectures.

**Introduction**

Hybrid superconductor–semiconductor devices have emerged as a promising platform for hosting topologically protected quantum states [1-3], which are essential for fault-tolerant quantum-computing architectures [4]. In these systems, quantum dots (QDs) coupled to superconductors (SCs) host correlative Andreev bound states (ABSs) [5-18], whose fermion parity, specifically whether even or odd, governs their low-energy electronic transport properties. Parity control in QD–SC systems has been achieved by tuning local parameters, such as the QD-SC [12,14,16] or QD-ABS [15,19] coupling strength, or by applying a magnetic field [12]. However, relying exclusively on local tuning and detection imposes significant integration challenges as the number of QDs increases. Furthermore, scaling these architectures towards artificial Kitaev chains [20-25] necessitates phase-control loops [26-29], which, while essential for topological tuning, can limit conventional local parity accessibility. Consequently, advancing these architectures demands non-local manipulation as a versatile tool to augment parity engineering. In parallel, supercurrent-based detection offers a scalable strategy to overcome the wiring complexity of conventional methods. Being intrinsically sensitive to the global parity, it provides a robust readout mechanism capable of accessing parities in extended QD–SC arrays where local access is restricted.

Recent experiments have demonstrated that Andreev molecules constitute an excellent platform for non-local manipulation [30-36], enabling access to characteristic non-local phenomena through phase or gate-voltage control, such as non-local tuning of the tunneling spectrum [36], the non-local Josephson effect [30-32,37,38], and the Josephson diode effect [33-35,39]. These studies also establish that the supercurrent is inherently sensitive to the non-local hybridization of ABSs, highlighting its potential as an intrinsic probe of non-local physics. Moreover, the capability for non-local manipulation is deeply rooted in the formation of Andreev molecule states via the hybridization of ABSs [37,39], realized in both Josephson junction (JJ)-based [30-37,39-42] and QD-related [19,38,43-46] device configurations. In particular, the combined QD-JJ-based Andreev molecules are uniquely advantageous, as they preserve the coherent transport properties of JJs while introducing Coulomb interactions. This interplay engenders a rich landscape of parity-dependent dynamics essential for extended Kitaev-chain building blocks [38,47]. However, despite these advances, a critical functionality gap persists. While existing research on Andreev molecules has predominantly focused on probing continuous spectral features or transport asymmetries, the realization of deterministic non-local control over discrete parity states and the validation of supercurrent-based parity detection as a sensor-free protocol have yet to be experimentally established.

In this work, we demonstrate deterministic non-local control and supercurrent-based detection of the parity configurations (PCs) in a QD-JJ-based Andreev molecule. We show that the PC of a target QD (QD1), whether in a purely even parity configuration (EPC) or an alternating even–odd parity configuration (EOPC), can be manipulated non-locally. This is achieved without any local modification, solely by electrostatically tuning the coherent hybridization with an adjacent QD (QD2). Our experimental investigation spans three distinct joint PC regimes within the elastic co-tunneling (ECT) limit, allowing us to map the operational conditions of this non-local manipulation. Corroborating these measurements with theoretical simulations that generate a global phase diagram, we uncover that the observed control follows deterministic selection rules. These rules are dictated by the system's joint PC, which determines the allowed transitions based on the fundamental symmetry difference between ECT and crossed Andreev reflection (CAR). Concurrently, we establish the supercurrent, directly signaled by a zero-bias conductance peak, as an intrinsic and sensor-free probe of the PCs. Unlike subgap conductance measured via normal leads, it encodes parity information through the global superconducting response of the entire device. Taken together, our results establish a controlled and extensible platform for non-local parity manipulation and detection in hybrid QD–SC systems. This capability offers a key building block for parity-sensitive superconducting architectures, providing a validated experimental strategy relevant to artificial Kitaev chains with phase-control loops.

## Results

### Device and measurement configuration

Figure 1(a) presents a scanning electron microscope (SEM) image of the reported device, with a schematic illustration shown in Fig. 1(b). The device consists of two QDs (QD1 and QD2) formed in an InAs nanowire proximitized by epitaxial aluminum, constituting a QD-JJ-based Andreev molecule. This geometry is designed to enable independent electrostatic control over each QD while maintaining the coherent superconducting coupling, which is essential for realizing non-local parity manipulation.

The InAs nanowires were grown by molecular beam epitaxy, followed by an *in-situ* epitaxial half-shell of 5 nm-thick Al deposited at low temperature to ensure a pristine superconductor–semiconductor interface [48]. QD2 was subsequently defined within the nanowire using three bottom finger gates after selectively wet-etching a 170 nm segment of the Al shell. An additional Al layer was then deposited to form superconducting leads, and a superconducting loop was incorporated around QD2 to fix the superconducting phase difference. In contrast, QD1 was naturally formed as a "shadow junction" during the growth process, utilizing the shadowing effect between the dense nanowires. This lithography-free approach eliminates the need for chemical etching, thereby preserving the structural integrity of the nanowire [49] and yielding high-quality tunneling characteristics, as demonstrated in previous studies of tunneling coherence times [50]. The complete SEM image of the device is provided in Supplemental Material (see Fig. S1 [51]).

The device was measured in a dilution refrigerator at a base temperature of ~ 40 mK. The measurement configuration is shown in Fig. 1(b). The middle superconducting lead was always grounded, while the left superconducting lead was used to apply both an AC excitation voltage $V_{\text{ac}}$ and a DC bias voltage $V_{\text{bias}}$. This setup allows us to measure the AC current $I_{\text{ac}}$ and derive the local differential conductance $dI/dV$ of QD1. The chemical potentials of QD1 and QD2 were independently controlled by gate voltages $V_{\text{QD1}}$ and $V_{\text{QD2}}$, respectively. Furthermore, the tunnel couplings between QD2 and its adjacent superconducting hybrid segments were modulated by gate voltages $V_{\text{T1}}$ and $V_{\text{T2}}$. Further details regarding device fabrication and measurement setups are provided in the Supplemental Material [51].

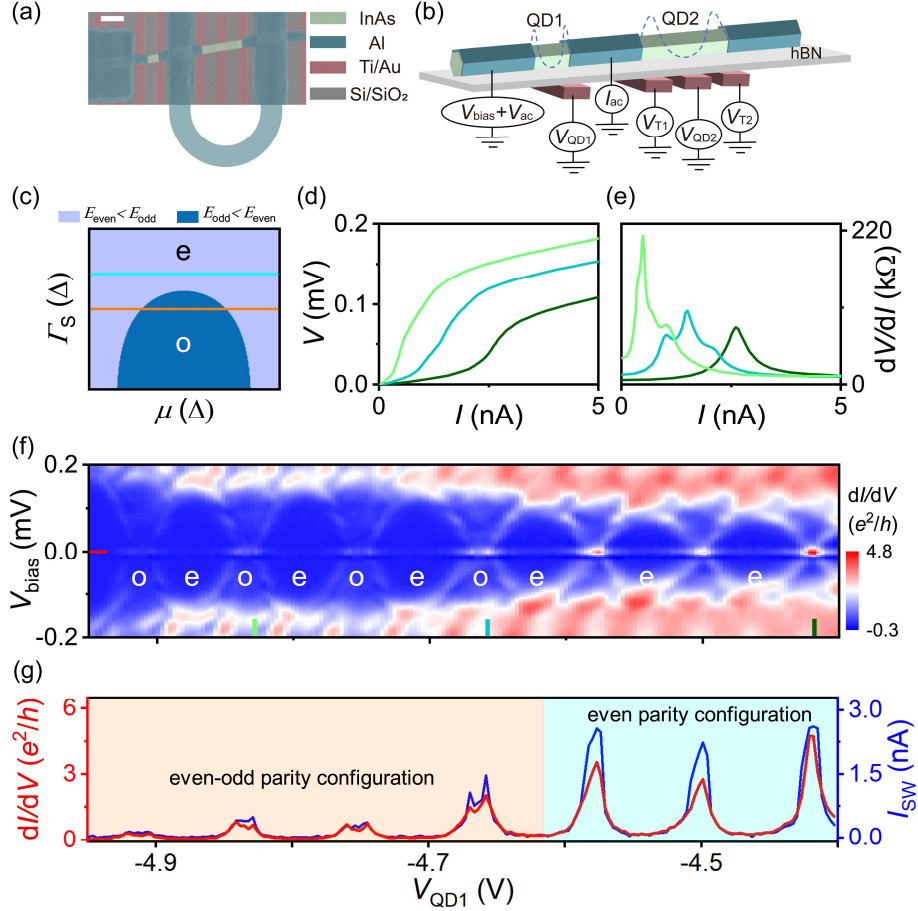

**FIG. 1. Local control of the parity of QD1.** (a) False-color SEM image of the device. The scale bar is 100 nm. The hBN dielectric layer between the Ti/Au finger gates and the nanowire is transparent in this image. The additional arc schematically indicates the superconducting loop, which is not visible in the SEM image. (b) Illustration of a QD JJ-based Andreev molecule and the measurement configuration. The dashed lines denote natural or gate-defined tunnel barriers. (c) Phase diagram of a QD-SC system, illustrating the GS parity as a function of the QD–SC coupling strength $\Gamma_S$, the superconducting pairing potential $\Delta$, and the QD chemical potential $\mu$. (d) The $I$-$V$ curves, obtained through integration, for representative values of $V_{QD1}$ indicated by the vertical bars in panel (f). (e) Differential resistance $dV/dI$ as a function of $I$, obtained by numerical differentiation of the $I$-$V$ curves in (d). (f) Differential conductance $dI/dV$ as a function of $V_{bias}$ and $V_{QD1}$, measured at $V_{QD2} = V_{T1} = V_{T2} = 0$ V. The letters e and o represent the even and odd occupancy states, respectively. (g) Line cut of the zero-bias conductance along the horizontal red bar in panel (f) (red curve), together with the extracted switching supercurrent of QD1 as a function of $V_{QD1}$ (blue curve). The cyan (orange) shaded regions correspond to the even-parity (alternating even–odd parity) regimes, indicated by the cyan (orange) line in panel (c).

**Local control of the parity of QD1**

In a QD-SC system, the competition between superconductivity and Coulomb interactions governs the nature of its ground state (GS) and transport properties. Superconducting pairing favors even-occupancy, spin-singlet states, whereas Coulomb charging facilitates the stabilization of odd-occupancy states with a spin-doublet GS. The relative strength of the QD–SC coupling ($\Gamma_S$) determines whether the system resides in an even or odd parity sector [52], as summarized by the phase diagram shown in Fig. 1(c).

To experimentally demonstrate how coupling strength governs the GS, we tune the gate voltage $V_{\text{QD1}}$ and acquire the tunneling spectrum of QD1, as shown in Fig. 1(f). The evolution of the zero-bias conductance (ZBC) is detailed by a linecut (Fig. 1(g), red line) at the corresponding position indicated in Fig. 1(f). As $V_{\text{QD1}}$ is decreased, the ZBC evolves from a single peak into a split double-peak structure, signaling a transition between two distinct regimes, as shown in Fig. 1(g). In the single-peak regime (cyan region in Fig. 1(g)), the ABSs do not cross zero energy (see Fig. 1(f)), consistent with an EPC hosting a pure even GS. In contrast, the double-peak regime (orange region) exhibits zero-energy crossings of the ABSs (Fig. 1(f)), indicative of an EOPC including an odd-occupancy state. This spectral evolution directly reflects a reduction of the QD-SC coupling strength, which drives the system from the EPC into the EOPC. These experimental regimes are qualitatively consistent with the theoretical phases depicted by the cyan and orange lines in the phase diagram of Fig. 1(c).

Moreover, the single peak exhibits higher conductance compared to the double peaks, because enhanced $\Gamma_S$ favors Cooper pair tunneling [52]. To clarify the origin of the ZBC, the differential conductance data are integrated to yield the *I-V* curves. Figure 1(d) displays *I-V* curves for different values of $V_{\text{QD1}}$ indicated by the vertical bars in Fig. 1(f), showing only positive-current portions. The corresponding differential resistance (d*V*/d*I*) curves (Fig. 1(e)) are then obtained by numerical differentiation. Evident supercurrent signatures revealed in these curves indicate that conductance peaks originate from supercurrent contributions, despite a finite residual resistance. The d*V*/d*I* curves allow us to quantify the switching supercurrent from the position of the maximum peak. The resulting switching supercurrent as a function of $V_{\text{QD1}}$ is plotted as the blue curve in Fig. 1(g). Notably, the evolution of the supercurrent closely follows that of the ZBC across the parity transition. This strong correlation establishes that the ZBC features originate from supercurrent contributions and that the supercurrent provides a direct and sensitive probe of the local PC of QD1.

**Non-local control of the parity with QD1 in an EPC and QD2 in an EOPC**

Next, we examine the non-local parity control of QD1 by modulating the chemical potential of QD2. As emphasized in the introduction, the effectiveness of such non-local control depends critically on the joint PC of the two QDs. Our experimental investigation covers three distinct joint-PC regimes. We begin with the first regime, where QD1 is initialized in an EPC and QD2 in an EOPC (Regime I). This configuration serves as a crucial control experiment, establishing the baseline for conditional non-local parity modulation.

To initialize the state of QD2, we tune the coupling strength between QD2 and the SC lead via $V_{\text{T1}}$ and $V_{\text{T2}}$. By fixing $V_{\text{T1}} = V_{\text{T2}} = -1.39$ V, the conductance spectrum confirms that QD2 resides in an EOPC (see Fig. S2 [51]). Simultaneously, the ZBC of QD1 in this regime (Fig. 2(c)) is consistent with the single-peak regime (the cyan region) shown in Fig. 1(g), confirming that QD1 is in an EPC. As discussed in Fig. 1(g), the ZBC here actually reflects the evolution of the supercurrent. Figure 2(a) shows the evolution of the ZBC as a function of $V_{\text{QD1}}$ and $V_{\text{QD2}}$. Pronounced avoided crossings are observed, manifesting as arc-shaped discontinuities in the otherwise continuous vertical ZBC features. These characteristics constitute a hallmark of coherent inter-dot hybridization. We emphasize that while supercurrents in double-QD JJ devices can depend on charge occupation [53] or spin configurations [54,55], previous studies did not explore non-local effects nor report such clear hybridization signatures in the supercurrent response. Crucially, we rule out inter-gate capacitive crosstalk as the origin of these avoided crossings, as crosstalk would merely induce linear energy level shifts rather than the observed spectral splitting. Moreover, the effect of capacitive crosstalk is quantitatively negligible in our device (see Fig. S2 [51]). Notably, the magnitude of the avoided crossings, and thus the coherent hybridization strength, can be tuned by modifying the coupling between one QD and the superconducting lead (see Fig. S3 [51]).

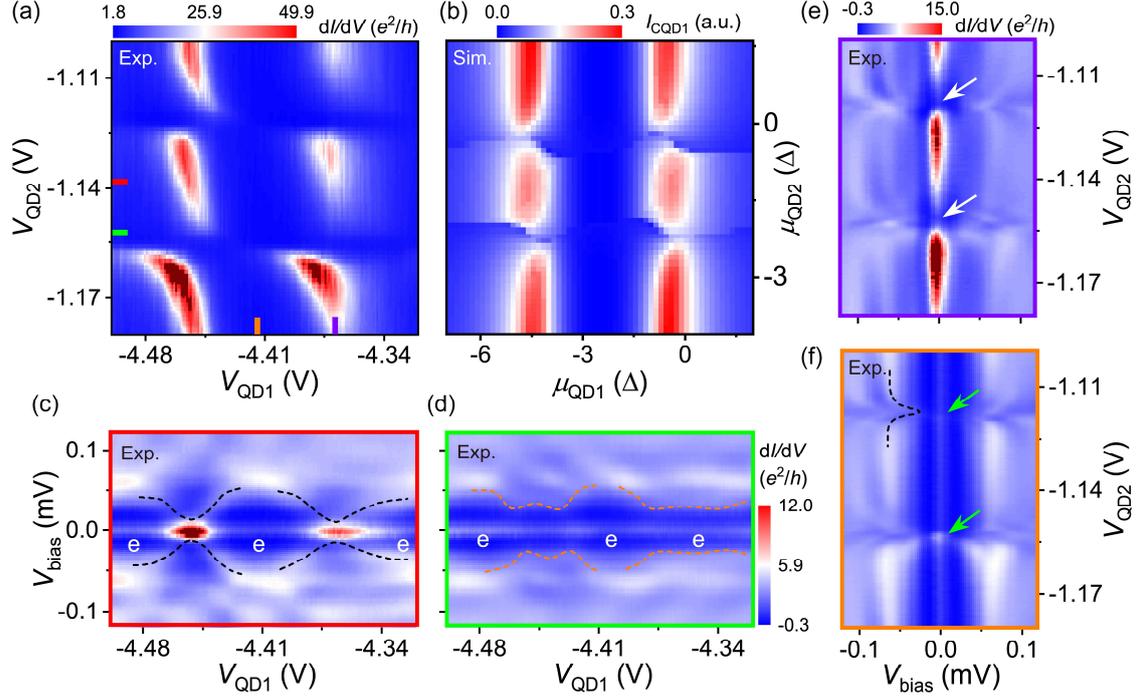

**FIG. 2. Non-local control of the parity with QD1 in an EPC and QD2 in an EOPC.** (a) $dI/dV$ as a function of $V_{QD1}$ and $V_{QD2}$, where $V_{bias} = 0$ mV and $V_{T1} = V_{T2} = -1.39$ V. (b) Simulated supercurrent of QD1 ($I_{CQD1}$), as a function of the QDs' chemical potentials $\mu_{QD1}$ and $\mu_{QD2}$. (c), (d) $dI/dV$ as a function of $V_{bias}$ and $V_{QD1}$, recorded at different values of $V_{QD2}$ indicated by the horizontal red and green bars in (a). The black and orange dashed lines guide the evolution of ABSs in (c), (d), respectively. (e), (f) $dI/dV$ as a function of $V_{bias}$ and $V_{QD2}$, recorded at different values of $V_{QD1}$ indicated by the vertical purple and orange bars in (a). The black dashed line in (f) guides the evolution of ABSs. Conductance values exceeding the color scale in (c) and (e) are displayed in wine red.

To elucidate the microscopic origin of these features, we simulate the supercurrent through QD1 as a function of the QDs' chemical potentials $\mu_{QD1}$ and $\mu_{QD2}$. To account for the EPC and EOPC for QD1 and QD2, respectively, the model incorporates a specific charge filling sequence, i.e., pairs for QD1 and single electrons for QD2 (see Fig. S4 [51]). Although a slight quantitative difference exists between the simulated supercurrent in Fig. 2(b) and the experimental conductance peaks in Fig. 2(a), the model qualitatively captures the essential avoided-crossing structure. These avoided crossings occur at values of $\mu_{QD2}$ where the charge number in QD2 changes, confirming that the non-local effect is mediated by the coherent hybridization with the adjacent QD's discrete levels. Details of the model are provided in Supplemental Material [51].

Further insight is gained by examining the tunneling spectra of QD1 (Figs. 2(c) and 2(d)) as a function of $V_{\text{QD1}}$ and $V_{\text{bias}}$ at fixed values of $V_{\text{QD2}}$, as indicated by the horizontal bars in Fig. 2(a). A comparison of the ABSs' evolution, guided by the black and orange dashed lines, reveals a marked difference: the ABSs in Fig. 2(d) are repelled away from zero energy. While such repulsion has previously been associated with the near-zero-energy ABSs hosted in the hybrid segment [45], in our device it likely originates from the coherent repulsion by the near-resonant ABSs of QD2. Importantly, regarding the PC, despite these non-local spectral modifications, the ABSs in QD1 are detuned far from zero energy (Fig. 2(d)), and no PC transition is induced—QD1 remains in the EPC. These results demonstrate that, in this specific joint PC, non-local control does not alter the PC of QD1, validating the first entry of our selection rules (Table I). This regime therefore serves as a control experiment, demonstrating that non-local coupling alone is insufficient to induce a PC transition without the appropriate joint PC. It is worth noting that, although we focus here on probing the parity of QD1, the supercurrent is governed by the total ground-state energy of the molecule, which includes contributions from both QD1 and QD2. Therefore, the supercurrent is intrinsically sensitive to the parity of QD2 as well, effectively serving as a probe for the total parity.

We next analyze the effect of tuning QD2 on the spectrum of QD1 at fixed values of $V_{\text{QD1}}$, corresponding to the vertical bars in Fig. 2(a). The resulting spectra are shown in Figs. 2(e) and 2(f). At the $V_{\text{QD1}}$ position marked by the purple bar, the ABSs in QD1 are tuned to (nearly) zero energy. As shown in Fig. 2(e), when QD2 is brought on resonance, the supercurrent conductance peak is strongly suppressed, as indicated by the white arrows. We attribute this suppression to the reduced spectral weight of the near-zero-energy ABSs in QD1 due to the hybridization with QD2. In contrast, at the $V_{\text{QD1}}$ position marked by the orange bar, where the ABSs in QD1 are tuned far from zero energy (Fig. 2(f)), tuning QD2 on resonance produces a qualitatively different response. The finite-bias ABSs shift toward zero energy (black dashed line), resulting in a modest enhancement of the supercurrent conductance peak, as indicated by the green arrows. This distinct behavior highlights how the non-local effect of QD2 critically depends on the energetic position of the ABSs in QD1.

**Non-local control of the parity with both QD1 and QD2 in an EOPC**

We now turn to the second joint-PC regime, where both QD1 and QD2 reside in an EOPC (Regime II). This is the crucial regime where deterministic non-local parity control is enabled.

We first demonstrate the realization of the EOPC in QD1. This can be achieved via two distinct mechanisms, i.e., by applying a magnetic field or reducing the coupling strength between QD1 and the SC leads. Figure 3(a) shows the zero-bias conductance map measured at a magnetic field of $B = 80$ mT, which contrasts sharply with the zero-field measurement shown in Fig. 2(a) for the same ($V_{\text{QD1}}$, $V_{\text{QD2}}$) parameter range, where QD1 exhibits an EPC. This change reflects a magnetic-field-induced quantum phase transition, as illustrated by the calculated phase diagram in Fig. 3(d). At zero magnetic field (black dashed line in Fig. 3(d), corresponding to Fig. 2(a)), the GS is purely even; however, under a finite magnetic field (orange dashed line in Fig. 3(d), corresponding to Fig. 3(a)), the GS includes odd states. Mechanistically, when the QD's GS is an even state and the excited state is an odd state, a magnetic field splits the spin-degenerate odd state, leading to a quantum phase transition that transforms the GS from even to odd [12]. Experimentally, this transition manifests as a bifurcation of the single conductance peak in Fig. 2(a) into the split double-peak structure in Fig. 3(a). This is consistent with the phase diagram shown in Fig. 3(d), confirming the switch of QD1 from EPC to EOPC.

Alternatively, we can access the EOPC regime by reducing the coupling strength between QD1 and the SC leads. This is achieved by sweeping $V_{\text{QD1}}$ to a more negative range and measuring conductance maps at both zero and finite $B$. The conductance map in Fig. 3(b) is measured at $B = 80$ mT with a more negative $V_{\text{QD1}}$ than in Fig. 3(a), whereas the zero-field map for this gate setting is displayed in Supplemental Fig. S5(a) [51]. Compared to the EPC behavior in Fig. 2(c), the spectrum in Fig. S5(b) clearly exhibits an EOPC for this more negative $V_{\text{QD1}}$ range [51], consistent with the behavior established in Fig. 1(f), where reducing $\Gamma_\text{S}$ can drive a PC transition from EPC to EOPC. This gate-induced transition is a distinct mechanism from the magnetic-field-induced transition shown in Fig. 3(a). Figure 3(e) illustrates the corresponding evolution of the phase diagram with magnetic field, where the QD is already in an EOPC at zero field. In this case, the GS includes an odd state, and the application of a magnetic field merely splits the odd state but does not induce a quantum phase transition [12], such that the PC remains unchanged. Accordingly, the black and cyan dashed lines in Fig. 3(e) correspond to Fig. S5(a) (zero field) and Fig. 3(b) (finite field), respectively, with QD1 in an EOPC in both cases. Meanwhile, the behavior of the conductance peak in Fig. 3(b) closely resembles the evolution observed in one half of the parameter space of Fig. 3(a). These results confirm that the PC of QD1 is qualitatively identical in Figs. 3(a) and (b), with both QD1 and QD2 residing in an EOPC.

The charge stability diagrams observed in Figs. 3(a) and 3(b) closely resemble those reported in studies of Kitaev chains [21-24,27,28], where the orientation of the avoided-crossing features serves as a diagnostic of the dominant inter-dot coupling mechanism. In Fig. 3(a), the presence of clear avoided crossings signifies a strong coupling between the two QDs, whereas their anti-diagonal orientation reveals that ECT dominates the inter-dot coupling. Compared with Fig. 3(a), the avoided crossings in Fig. 3(b) are even more pronounced, and the conductance peaks along the horizontal orientation are almost invisible. The latter stems from the fact that the measurement probes the local conductance of QD1. Despite these subtle differences, the overall behavior of the conductance peaks in Fig. 3(b) remains qualitatively consistent with that in Fig. 3(a). Figure 3(c) presents the double-QD phase diagram obtained from a theoretical simulation corresponding to the data in Fig. 3(b), where ECT is set to dominate the coupling. The simulation reproduces the experimental phase boundaries of Fig. 3(b), reinforcing our identification of the dominant coupling mechanism (see Supplemental Material [51] for details).

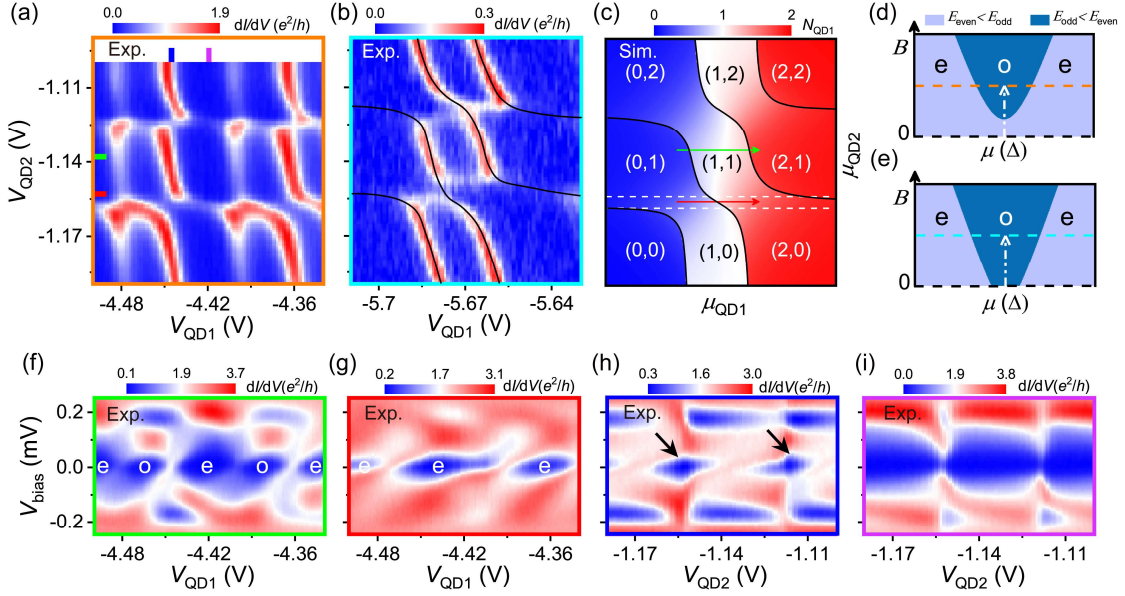

**FIG. 3. Non-local control of the parity with both QD1 and QD2 in an EOPC.** (a) $dI/dV$ as a function of $V_{QD1}$ and $V_{QD2}$, where $V_{bias} = 0$ mV, $V_{T1} = V_{T2} = -1.39$ V and $B = 80$ mT. The white region for $V_{QD2} > -1.1$ V contains no data points. (b) $dI/dV$ as a function of $V_{QD1}$ and $V_{QD2}$, measured under the same conditions as in (a) but for a more negative $V_{QD1}$ range, corresponding to a reduced QD1–SC coupling strength. (c) Simulated charge stability diagram of a double-QD system. The color scale indicates the charge number of QD1 as a function of QDs' chemical potentials $\mu_{QD1}$ and $\mu_{QD2}$. The solid black lines denote the even–odd charge-state boundaries. The texts indicate the GS charge occupation numbers of a double QD system. (d), (e) Evolution of the GS of the QD-SC system as a function of $B$ and chemical potential $\mu$. At zero magnetic field, the QD in (d) hosts a purely even GS, whereas it includes an odd GS in (e). The orange and cyan dashed lines correspond to (a) and (b), respectively. White arrows indicate the evolution from zero to finite magnetic field. (f), (g) $dI/dV$ as a function of $V_{bias}$ and $V_{QD1}$, recorded at different values of $V_{QD2}$ indicated by the horizontal green and red bars in (a). (h), (i) $dI/dV$ as a function of $V_{bias}$ and $V_{QD2}$, recorded at fixed values of $V_{QD1}$ indicated by the vertical blue and purple bars in (a).

Having established the regime and mechanism, we now demonstrate the non-local control of the parity in this second joint-PC regime. We focus on the response of QD1 as QD2 is tuned across its resonance. Figures 3(f) and 3(g) show the spectra of QD1 as a function of $V_{QD1}$ and $V_{bias}$ at fixed values of $V_{QD2}$, as indicated by the horizontal bars in Fig. 3(a). When QD2 is detuned (green bar, off-resonance), QD1 retains its characteristic EOPC double-peak structure (Fig. 3(f)). When QD2 is tuned to resonance (red bar), a striking transformation occurs: the double peaks merge into a single zero-bias peak, signifying that QD1 has transitioned to a purely EPC (Fig. 3(g)). A similar transition is also observed in the measurements corresponding to Fig. 3(b) (see Fig. S6 [51]). Taken together, these results demonstrate that when QD2 is tuned into resonance, non-local coupling induces a PC transition in QD1, provided that QD1 initially resides in an EOPC—whether the EOPC is realized by applying a magnetic field or by reducing the QD–SC coupling strength via gate tuning. This behavior contrasts sharply with the first joint-PC regime (Fig. 2), where QD1 remains in an EPC and no PC transition is induced, validating the conditional selectivity (Table I). Notably, the observed PC transition here necessarily involves a non-local parity transition of QD1 itself, as the conversion from an EOPC to a purely EPC corresponds to a change from odd to even parity driven by tuning the adjacent dot.

The physical mechanism driving this switch is elucidated by the theoretical phase diagram in Fig. 3(c). As shown in the diagram, strong coherent hybridization between the QDs opens a large avoided crossing gap when QD2 is on resonance. This creates a distinct energy window, bounded by the white dotted lines, which corresponds to the formation of a molecular bonding state [30,36,37]. Within this window, the bonding state (associated with the even-parity (2, 0) configuration) is energetically favored, thereby suppressing the odd-parity sector. Consequently, when sweeping $V_{\text{QD1}}$ along the resonant path (red arrow), the system transitions directly from the (0, 1) state to the (2, 0) state, effectively "skipping" the odd-parity regime. Under such conditions, QD1 manifests locally as a pure EPC. In contrast, sweeping away from the resonance of QD2 (green arrow) reduces the hybridization, restoring the standard even–odd filling sequence and the associated PCs. This mechanism explains why performing horizontal scans in Fig. 3(a) yields fundamentally different PCs in Fig. 3(f) (off-resonance) versus Fig. 3(g) (on-resonance).

To complement the picture, we investigate the response of the QD1 spectrum to sweeping $V_{\text{QD2}}$ at fixed values of $V_{\text{QD1}}$, corresponding to the vertical bars in Fig. 3(a). The resulting spectra are shown in Figs. 3(h) and 3(i). A clear distinction emerges: the ABSs of QD1 are displaced far from zero energy in Fig. 3(i), while they remain close to zero energy in Fig. 3(h). Notably, as highlighted by the black arrows in Fig. 3(h), avoided crossings of the ABSs occur precisely at zero energy. Analogous to this characteristic feature observed in systems where ABSs strongly hybridize with Yu-Shiba-Rusinov (YSR) states in a QD [45], this further provides spectroscopic evidence for the formation of the Andreev molecule states due to coherent hybridization that mediates the observed parity switch.

**Non-local control of the parity with QD1 in an EOPC and QD2 in an EPC**

Next, we examine the third regime of the joint PC, where QD1 is in an EOPC and QD2 is in an EPC (Regime III). This configuration is achieved by relaxing the tunnel-gate voltages to $V_{T1} = V_{T2} = 0$ V, which enhances the coupling between QD2 and the SC leads. Figures 4(a)-(c) show the zero-field spectra of QD1 as a function of $V_{QD1}$ and $V_{bias}$ at fixed values of $V_{QD2} = 0.09$ V, $0.112$ V, and $0.127$ V, respectively. The chosen $V_{QD1}$ range corresponds to the EOPC region of QD1, as identified in Figs. 1(f) and 1(g). The evolution of the corresponding ZBC is plotted in Fig. 4(d). As discussed in Fig. 1(g), the behavior of the ZBC directly reflects the evolution of the supercurrent and thus the PC of QD1. Specifically, the red and purple curves exhibit a double-peak structure, indicating that QD1 exhibits an EOPC in the regimes of Figs. 4(a) and 4(c). In contrast, the cyan curve displays a single peak, signifying that QD1 transitions into an EPC in the regime of Fig. 4(b). This behavior demonstrates that tuning $V_{QD2}$ and hence the chemical potential of the ABSs in QD2 can induce a PC transition in QD1 through non-local coupling. As in the second joint-PC regime, the PC transition observed here is accompanied by a non-local parity flip of QD1, from odd to even. We further note that the finite-bias ABS features in Fig. 4(b) appear distorted compared to those in Figs. 4(a) and 4(c). Such distortions are signatures of hybridization between ABSs in the two QDs, consistent with previous observations in other Andreev molecule systems [40,43].

To theoretically validate the PC transition observed in Figs. 4(a)-(c), we simulate the system in the ECT-dominant regime where QD1 starts in an EOPC and QD2 in an EPC, resulting in the phase diagram presented in Fig. 4(e). While similar phase diagrams have been observed for a QD strongly coupled to singlet YSR states or ABSs employing direct tunnel coupling [15,19], our Andreev molecule system incorporates two distinct inter-dot coupling channels, i.e., a spin-conserving tunneling process and a spin-flipping tunneling process (see Supplemental Material [51] for details). The colored horizontal bars in Fig. 4(e) correspond to the experimental conditions of Figs. 4(a)-(c). Along the green bar, QD1 resides in an EOPC; along the orange bar, QD1 transitions into an EPC; and along the blue bar, the EOPC re-emerges. In the simulation, QD2 is maintained in an EPC, demonstrating that the enhanced coupling between QD2 and the SC leads causes the PC of QD1 to transition from EOPC to EPC. These results demonstrate that the PC transition in QD1 observed in Fig. 4(b) is driven by non-local control, and can be identified through the behavior of the supercurrent conductance peaks.

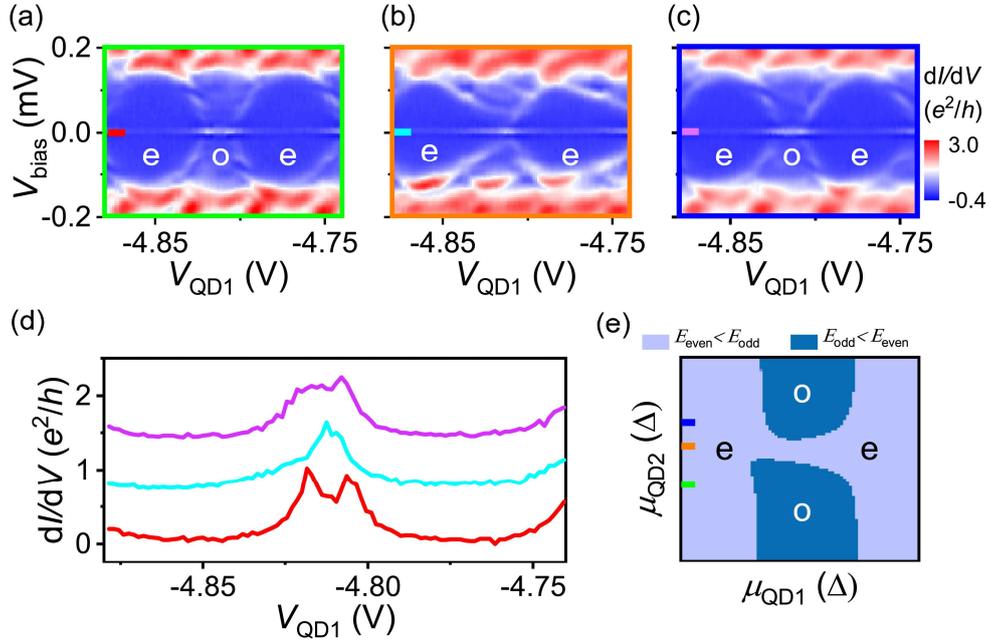

**FIG. 4. Non-local control of the parity with QD1 in an EOPC and QD2 in an EPC.** (a), (b), (c) $dI/dV$ as a function of $V_{bias}$ and $V_{QD1}$, with $V_{T1} = V_{T2} = 0\,\text{V}$ and $B = 0\,\text{mT}$, recorded at $V_{QD2} = 0.09\,\text{V}$, $0.112\,\text{V}$, and $0.127\,\text{V}$, respectively. (d) Zero-bias line cuts, with line colors corresponding to the horizontal bars in (a)-(c). For clarity, a vertical offset is applied. (e) Simulated phase diagram of the Andreev-molecule system corresponding to the experimental conditions in (a)–(c).

Taken together, our measurements in the ECT-dominant regime establish a set of clear selection rules for non-local parity control in the Andreev molecule: 1) tuning QD2 can induce a non-local PC (and hence parity) transition of QD1 only when QD1 initially resides in an EOPC; 2) no transition is observed when QD1 is in a purely EPC (Table I). To further generalize the physical principles underlying the observed PC transitions, we performed extensive parameter-space sampling based on our theoretical model, with the results summarized in Table SI [51]. These simulations confirm that, when ECT dominates the inter-dot coupling, the selection rules discussed above are robust—fully consistent with the experimental observations (Table I). In contrast, in the CAR-dominant limit (not accessed in the present experiment), the model predicts a modified set of selection rules. A PC transition is allowed even when QD1 is in an EPC, whereas it is forbidden when QD1 is in an EOPC and QD2 remains in an EPC. We believe that the microscopic origin of these diverging selection rules lies in the distinct symmetry of the coupling mechanisms. ECT conserves the total electron number on the dots, enabling parity transitions only by redistributing existing quasiparticles (e.g., $(1, 1) \leftrightarrow (2, 0)$). Conversely, CAR involves the creation or annihilation of pairs from the superconducting condensate, thereby enabling transitions by directly coupling the vacuum to the two-electron sector (e.g., $(0, 0) \leftrightarrow (1, 1)$). Collectively, these established universal selection rules show that, beyond the joint PC of the coupled QDs, the dominant inter-dot coupling channel (ECT vs. CAR) provides an additional knob that qualitatively dictates the outcome of non-local parity control.

**Conclusion and discussion**

To conclude, we have demonstrated deterministic non-local control of PC transitions and established the supercurrent as an intrinsic probe of parity in a QD-JJ-based Andreev molecule. To determine the principles governing this control, we systematically investigated three distinct joint-PC regimes within the ECT-dominant limit. Our measurements reveal that the success of non-local control is conditional. A non-local PC transition is observed in Regimes II and III (where the target QD resides in an EOPC), but is forbidden in Regime I (where it resides in an EPC). Synthesizing these experimental observations with theoretical simulations, which extend the analysis to the CAR-dominant limit, we established a set of universal selection rules. Crucially, these rules highlight that the outcome of non-local control is strictly dictated by the symmetry-imposed interplay between the joint PC of the system and the dominant inter-dot coupling mechanism. In principle, this coupling can be tuned by adjusting the chemical potential of the intermediate hybrid segment [56], enabling the location of the "sweet spot" [21,23]. Future extensions that incorporate additional tunability of the intermediate hybrid segment are expected to enable the realization of longer topological Kitaev chains.

The implications of this work extend beyond basic parity engineering. Our results demonstrate a powerful, sensor-free strategy for managing and detecting parity in multi-QD architectures, critical for future artificial Kitaev chains incorporating phase-control loops [26,28,29]. By obviating the need for auxiliary sensors, supercurrent-based detection alleviates wiring bottlenecks while naturally suppressing quasiparticle poisoning compared to normal-metal probes. Finally, although currently operating in the low-frequency limit, this method's inherent compatibility with circuit-QED architectures facilitates straightforward integration with fast readout techniques [57,58]. These combined advantages position Andreev-molecule-based architectures as a promising building block for scalable, parity-controlled superconducting quantum devices.

**Table** I**. Summary of non-local control of the PC of QD1**

| Regimes | Initial PC of QD1 | Initial PC of QD2 | Initialization Strategy (Realization of joint PC) | Outcome of non-local control | Key experimental signatures (QD2 on resonance) |
|---|---|---|---|---|---|
| Regime I (Fig. 2) | EPC | EOPC | Strong QD1–SC coupling; Weak QD2–SC coupling | No PC transition | Avoided crossings of supercurrent |
| Regime II (Fig. 3) | EOPC | EOPC | Magnetic field induced phase transition OR Reduced QD1–SC coupling | PC transition (including QD1 parity flip) | Avoided crossings of ABSs |
| Regime III (Fig. 4) | EOPC | EPC | Reduced QD1–SC coupling; Strong QD2–SC coupling | PC transition (including QD1 parity flip) | ZBC evolution from split to single peak |

Note: PC, parity configuration; QD, quantum dot; EPC, even parity configuration; EOPC, even–odd parity configuration; SCs, superconductors; ABSs, Andreev bound states; ZBC, zero-bias conductance.

**Acknowledgments**

This work was supported by the National Key Research and Development Program of China (2022YFA1403400 and 2025YFA1411400); by the National Natural Science Foundation of China (12074417, 92065203, 92365207, 12374459, 61974138, and 92065106); by the Strategic Priority Research Program of Chinese Academy of Sciences (XDB33000000); by the Synergetic Extreme Condition User Facility sponsored by the National Development and Reform Commission; and by the Innovation Program for Quantum Science and Technology (2021ZD0302600, 2021ZD0301800 and 2021ZD0302400). D. P. acknowledges the support from the State Key Laboratory of Micro-nano Engineering Science (MES202601) and Youth Innovation Promotion Association, Chinese Academy of Sciences (2017156 and Y2021043).

**Data Availability**

The data that support the findings of this article will be openly available [link to be provided].

**References**

[1] R. M. Lutchyn, J. D. Sau, and S. Das Sarma, Majorana Fermions and a Topological Phase Transition in Semiconductor-Superconductor Heterostructures. Physical Review Letters **105**, 077001 (2010).

[2] R. M. Lutchyn, E. P. A. M. Bakkers, L. P. Kouwenhoven, P. Krogstrup, C. M. Marcus, and Y. Oreg, Majorana zero modes in superconductor–semiconductor heterostructures. Nature Reviews Materials **3**, 52 (2018).

[3] E. Prada et al., From Andreev to Majorana bound states in hybrid superconductor–semiconductor nanowires. Nature Reviews Physics **2**, 575 (2020).

[4] C. Nayak, S. H. Simon, A. Stern, M. Freedman, and S. Das Sarma, Non-Abelian anyons and topological quantum computation. Reviews of Modern Physics **80**, 1083 (2008).

[5] M. R. Buitelaar, T. Nussbaumer, and C. Schönenberger, Quantum Dot in the Kondo Regime Coupled to Superconductors. Physical Review Letters **89**, 256801 (2002).

[6] A. Eichler, M. Weiss, S. Oberholzer, C. Schönenberger, A. Levy Yeyati, J. C. Cuevas, and A. Martín-Rodero, Even-Odd Effect in Andreev Transport through a Carbon Nanotube Quantum Dot. Physical Review Letters **99**, 126602 (2007).

[7] T. Sand-Jespersen et al., Kondo-Enhanced Andreev Tunneling in InAs Nanowire Quantum Dots. Physical Review Letters **99**, 126603 (2007).

[8] R. S. Deacon, Y. Tanaka, A. Oiwa, R. Sakano, K. Yoshida, K. Shibata, K. Hirakawa, and S. Tarucha, Tunneling Spectroscopy of Andreev Energy Levels in a Quantum Dot Coupled to a Superconductor. Physical Review Letters **104** (2010).

[9] J. D. Pillet, C. H. L. Quay, P. Morfin, C. Bena, A. L. Yeyati, and P. Joyez, Andreev bound states in supercurrent-carrying carbon nanotubes revealed. Nature Physics **6**, 965 (2010).

[10] E. J. H. Lee, X. Jiang, R. Aguado, G. Katsaros, C. M. Lieber, and S. De Franceschi, Zero-Bias Anomaly in a Nanowire Quantum Dot Coupled to Superconductors. Physical Review Letters **109**, 186802 (2012).

[11] R. Maurand, T. Meng, E. Bonet, S. Florens, L. Marty, and W. Wernsdorfer, First-Order 0−π Quantum Phase Transition in the Kondo Regime of a Superconducting Carbon-Nanotube Quantum Dot. Physical Review X **2**, 011009 (2012).

[12] E. J. H. Lee, X. Jiang, M. Houzet, R. Aguado, C. M. Lieber, and S. De Franceschi, Spin-resolved Andreev levels and parity crossings in hybrid superconductor–semiconductor nanostructures. Nature Nanotechnology **9**, 79 (2014).

[13] A. Jellinggaard, K. Grove-Rasmussen, M. H. Madsen, and J. Nygård, Tuning Yu-Shiba-Rusinov states in a quantum dot. Physical Review B **94**, 064520 (2016).

[14] S. Li, N. Kang, P. Caroff, and H. Q. Xu, 0−π phase transition in hybrid superconductor–InSb nanowire quantum dot devices. Physical Review B **95**, 014515 (2017).

[15] K. Grove-Rasmussen, G. Steffensen, A. Jellinggaard, M. H. Madsen, R. Žitko, J. Paaske, and J. Nygård, Yu–Shiba–Rusinov screening of spins in double quantum dots. Nature Communications **9**, 2376 (2018).

[16] J. C. Estrada Saldaña, A. Vekris, V. Sosnovtseva, T. Kanne, P. Krogstrup, K. Grove-Rasmussen, and J. Nygård, Temperature induced shifts of Yu–Shiba–Rusinov resonances in nanowire-based hybrid quantum dots. Communications Physics **3**, 125 (2020).

[17] Z. Scherübl et al., Large spatial extension of the zero-energy Yu–Shiba–Rusinov state in a magnetic field. Nature Communications **11**, 1834 (2020).

[18] J. He et al., Nonequilibrium interplay between Andreev bound states and Kondo effect. Physical Review B **102**, 075121 (2020).

[19] A. Bordin et al., Impact of Andreev Bound States within the Leads of a Quantum Dot Josephson Junction. Physical Review X **15**, 011046 (2025).

[20] A. Tsintzis, R. S. Souto, and M. Leijnse, Creating and detecting poor man's Majorana bound states in interacting quantum dots. Physical Review B **106** (2022).

[21] T. Dvir et al., Realization of a minimal Kitaev chain in coupled quantum dots. Nature **614**, 445 (2023).

[22] S. L. D. ten Haaf et al., A two-site Kitaev chain in a two-dimensional electron gas. Nature **630**, 329 (2024).

[23] F. Zatelli et al., Robust poor man's Majorana zero modes using Yu-Shiba-Rusinov states. Nature Communications **15**, 7933 (2024).

[24] I. Kulesh et al., Flux-Controlled Two-Site Kitaev Chain. Physical Review Letters **135**, 056301 (2025).

[25] Nick van Loo1 et al., Single-shot parity readout of a minimal Kitaev chain. arXiv:2507.01606v1 (2025).

[26] C.-X. Liu, S. Miles, A. Bordin, S. L. D. ten Haaf, G. P. Mazur, A. M. Bozkurt, and M. Wimmer, Scaling up a sign-ordered Kitaev chain without magnetic flux control. Physical Review Research **7** (2025).

[27] A. Bordin et al., Enhanced Majorana stability in a three-site Kitaev chain. Nature Nanotechnology **20**, 726 (2025).

[28] S. L. D. ten Haaf et al., Observation of edge and bulk states in a three-site Kitaev chain. Nature **641**, 890 (2025).

[29] Alberto Bordin et al., Probing Majorana localization of a phase-controlled three-site Kitaev chain with an additional quantum dot. arXiv:2504.13702v1 (2025).

[30] S. Matsuo, J. S. Lee, C.-Y. Chang, Y. Sato, K. Ueda, C. J. Palmstrøm, and S. Tarucha, Observation of nonlocal Josephson effect on double InAs nanowires. Communications Physics **5**, 221 (2022).

[31] D. Z. Haxell et al., Demonstration of the Nonlocal Josephson Effect in Andreev Molecules. Nano Letters **23**, 7532 (2023).

[32] S. Matsuo, T. Imoto, T. Yokoyama, Y. Sato, T. Lindemann, S. Gronin, G. C. Gardner, M. J. Manfra, and S. Tarucha, Phase engineering of anomalous Josephson effect derived from Andreev molecules. Science Advances **9**, eadj3698 (2023).

[33] S. Matsuo, T. Imoto, T. Yokoyama, Y. Sato, T. Lindemann, S. Gronin, G. C. Gardner, M. J. Manfra, and S. Tarucha, Josephson diode effect derived from short-range coherent coupling. Nature Physics **19**, 1636 (2023).

[34] S. Matsuo et al., Shapiro response of superconducting diode effect derived from Andreev molecules. Physical Review B **111**, 094512 (2025).

[35] S. Zhu et al., Josephson diode effect in nanowire-based Andreev molecules. Communications Physics **8**, 330 (2025).

[36] S. Matsuo et al., Phase-dependent Andreev molecules and superconducting gap closing in coherently-coupled Josephson junctions. Nature Communications **14**, 8271 (2023).

[37] J. D. Pillet, V. Benzoni, J. Griesmar, J. L. Smirr, and Ç. Ö. Girit, Nonlocal Josephson Effect in Andreev Molecules. Nano Letters **19**, 7138 (2019).

[38] M. Kocsis, Z. Scherübl, G. Fülöp, P. Makk, and S. Csonka, Strong nonlocal tuning of the current-phase relation of a quantum dot based Andreev molecule. Physical Review B **109**, 245133 (2024).

[39] J. D. Pillet, S. Annabi, A. Peugeot, H. Riechert, E. Arrighi, J. Griesmar, and L. Bretheau, Josephson diode effect in Andreev molecules. Physical Review Research **5**, 033199 (2023).

[40] M. Coraiola et al., Phase-engineering the Andreev band structure of a three-terminal Josephson junction. Nature Communications **14**, 6784 (2023).

[41] M. Coraiola et al., Flux-Tunable Josephson Diode Effect in a Hybrid Four-Terminal Josephson Junction. ACS Nano **18**, 9221 (2024).

[42] M. Coraiola et al., Spin-Degeneracy Breaking and Parity Transitions in Three-Terminal Josephson Junctions. Physical Review X **14**, 031024 (2024).

[43] O. Kürtössy, Z. Scherübl, G. Fülöp, I. E. Lukács, T. Kanne, J. Nygård, P. Makk, and S. Csonka, Andreev Molecule in Parallel InAs Nanowires. Nano Letters **21**, 7929 (2021).

[44] O. Kürtössy et al., Heteroatomic Andreev molecule in a superconducting island-double quantum dot. arXiv:2407.00825 (2024).

[45] D. van Driel et al., Charge Sensing the Parity of an Andreev Molecule. PRX Quantum **5**, 020301 (2024).

[46] Z. Su, A. B. Tacla, M. Hocevar, D. Car, S. R. Plissard, E. P. A. M. Bakkers, A. J. Daley, D. Pekker, and S. M. Frolov, Andreev molecules in semiconductor nanowire double quantum dots. Nature Communications **8**, 585 (2017).

[47] J. D. Sau and S. D. Sarma, Realizing a robust practical Majorana chain in a quantum-dot-superconductor linear array. Nature Communications **3**, 964 (2012).

[48] D. Pan et al., In Situ Epitaxy of Pure Phase Ultra-Thin InAs-Al Nanowires for Quantum Devices. Chinese Physics Letters **39**, 058101 (2022).

[49] S. A. Khan et al., Highly Transparent Gatable Superconducting Shadow Junctions. ACS Nano **14**, 14605 (2020).

[50] J. He et al., Quantifying quantum coherence of multiple-charge states in tunable Josephson junctions. npj Quantum Information **10**, 1 (2024).

[51] See Supplemental Material for additional supporting data and complete theoretical models.

[52] S. De Franceschi, L. Kouwenhoven, C. Schönenberger, and W. Wernsdorfer, Hybrid superconductor–quantum dot devices. Nature Nanotechnology **5**, 703 (2010).

[53] J. C. Estrada Saldaña, A. Vekris, G. Steffensen, R. Žitko, P. Krogstrup, J. Paaske, K. Grove-Rasmussen, and J. Nygård, Supercurrent in a Double Quantum Dot. Physical Review Letters **121**, 257701 (2018).

[54] D. Bouman et al., Triplet-blockaded Josephson supercurrent in double quantum dots. Physical Review B **102**, 220505(R) (2020).

[55] R. Debbarma, M. Aspegren, and F. V. Boström, Josephson current via spin and orbital states of a tunable double quantum dot. Physical Review B **106**, LI80507 (2022).

[56] A. Bordin et al., Tunable Crossed Andreev Reflection and Elastic Cotunneling in Hybrid Nanowires. Physical Review X **13**, 031031 (2023).

[57] X. Yang et al., Procedure to tune a three-site artificial Kitaev chain to host Majorana bound states based on transmon measurements. Physical Review B **112**, 165418 (2025).

[58] Enna Zhuo et al., Measurement of parity-dependent energy-phase relation of the low-energy states in a potential artificial Kitaev chain utilizing a transmon qubit. arXiv:2501.13367v2 (2025).

# Supplemental Material for Deterministic non-local parity control and supercurrent-based detection in an Andreev molecule

Shang Zhu[1,2,#], Xiaozhou Yang[1,2,#], Mingli Liu[1], Min Wei[1,2], Yiping Jiao[1,2], Jiezhong He[1,2], Bingbing Tong[1,3], Junya Feng[1], Ziwei Dou[1], Peiling Li[1,3], Jie Shen[1], Xiaohui Song[1,3], Guangtong Liu[1,3], Zhaozheng Lyu[1,3,*], Dong Pan[4,*], Jianhua Zhao[5], Li Lu[1,2,3,*], Fanming Qu[1,2,3,*]

[1] Beijing National Laboratory for Condensed Matter Physics, Institute of Physics, Chinese Academy of Sciences, Beijing 100190, China
[2] University of Chinese Academy of Sciences, Beijing 100049, China
[3] Hefei National Laboratory, Hefei 230088, China
[4] State Key Laboratory of Semiconductor Physics and Chip Technologies, Institute of Semiconductors, Chinese Academy of Sciences, Beijing 100083, China
[5] State Key Laboratory of Spintronics, Hangzhou International Innovation Institute, Beihang University, Hangzhou 311115, China
[#] These authors contributed equally to this work.
[*] Email: lyuzhzh@iphy.ac.cn; pandong@semi.ac.cn; lilu@iphy.ac.cn; fanmingqu@iphy.ac.cn

## I. Supplemental methods

### A. Device fabrication

First, Ti/Au (3/10 nm) bottom finger gates were patterned on a Si/SiO$_2$ substrate, followed by the transfer of a ~30-nm-thick hBN flake onto the gates to serve as a dielectric layer. InAs nanowires were then transferred onto the hBN. Both the InAs nanowire and hBN were precisely positioned via a high-precision transfer system equipped with a hot-release polymer. This system operates by picking up materials at low temperature and releasing them at elevated temperature. The epitaxial aluminum shell of the InAs nanowires was selectively etched using Transene D, after which etching residues were removed using deionized water. Finally, Al electrodes (80 nm thick) were deposited after in-situ Ar$^+$ etching to remove the native oxide layer.

### B. Transport measurements

The transport measurements were carried out in a dilution refrigerator at ~40 mK. Differential conductance d$I$/d$V$ was measured using a standard low-frequency (17.77 Hz) lock-in technique (LI5650). A Keithley 2612 source meter was used to apply the DC bias voltage. The original data, i.e., the $dI/dV - V_{\text{bias}}$ curves, were acquired directly, while the $I - V$ curves (e.g., Fig. 1d) were obtained by numerical integration. All gate voltages were applied via the AUX channels of the LI5650 and a Keithley 2400 source meter, with a 10 MΩ resistor connected in series for circuit protection.

## II. Supplemental data

### A. Complete SEM image of the device

The complete SEM image of the device is provided in Fig. S1, while Fig. 1(a) shows a magnified view highlighting the central device features. A superconducting loop was incorporated around QD2 to fix the superconducting phase difference across the junction.

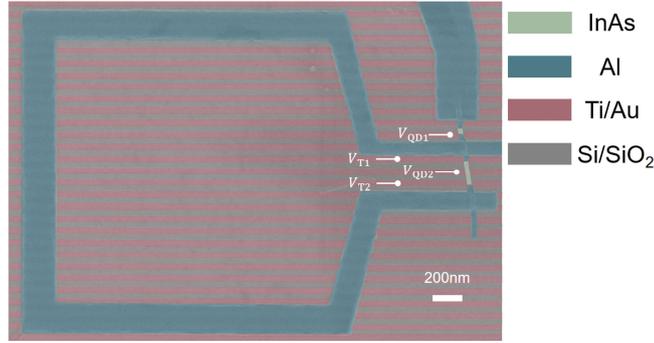

**FIG. S1. Complete SEM image of the device.** The white markers and associated labels identify the electrostatic gates used for electrostatic control.

### B. EOPC in QD2

As outlined in the main text, the first regime of the joint PC we investigate corresponds to QD1 in an EPC and QD2 in an EOPC. We now characterize the PC of QD2. Based on the evolution of the local conductance of QD1 under the modulation of three gates of QD2 ($V_{\text{T1}}$, $V_{\text{T2}}$ and $V_{\text{QD2}}$), we identify that QD2 is in an EOPC. First, the local conductance ($dI/dV$) of QD1 was measured as a function of $V_{\text{QD2}}$ and $V_{\text{T1}}$&$V_{\text{T2}}$ (Fig. S2(a)). Upon tuning $V_{\text{T1}}$ and $V_{\text{T2}}$, the conductance peak evolves from a single peak into a split double-peak structure, indicating that the PC of QD2 transitions from an EPC to an EOPC. With $V_{\text{T1}}$ and $V_{\text{T2}}$ fixed at the values indicated by the horizontal red bar in Fig. S2(a), the charge stability diagram (Fig. S2(b)) was obtained by measuring $dI/dV$ as a function of $V_{\text{QD1}}$ and $V_{\text{QD2}}$. This charge stability diagram exhibits pronounced avoided crossings, reflecting strong coherent coupling between the two QDs, and reveals an even–odd alternation of the charge parity of QD2 as a function of $V_{QD2}$. These observations demonstrate that QD2 is in an EOPC.

Next, we discuss the impact of inter-gate capacitive crosstalk on the experimental results. The slanted line in Fig. S2(a) indicates capacitive crosstalk between $V_{T1}\&V_{T2}$ and $V_{QD2}$. In the main text, the sweeping range of $V_{QD2}$ is restricted to -1.1 V and -1.18 V at $V_{T1}\&V_{T2} = -1.39$ V; within this window, the EOPC formed in QD2 remains relatively robust and unaffected by the crosstalk. There is also capacitive crosstalk between $V_{QD1}$ and $V_{QD2}$ (visible in Fig. S2(b)). Assuming a lever arm α from $V_{QD2}$ to $V_{QD1}$, compensation during non-local sweeps is applied following the relation $V'_{QD1} = V_{QD1} + \alpha V_{QD2}$. From the slope in Fig. S2(b), we estimate $\alpha \approx 0.1$. However, considering the narrow sweep range of $V_{QD2}$ (only 0.08 V), the induced voltage shift on $V_{QD1}$ is only ~0.008 V. This is significantly smaller than the voltage scale required for a parity transition in QD1 (Fig. 1(f)). Therefore, the crosstalk does not qualitatively affect our physical interpretation, and no numerical compensation was applied to the data.

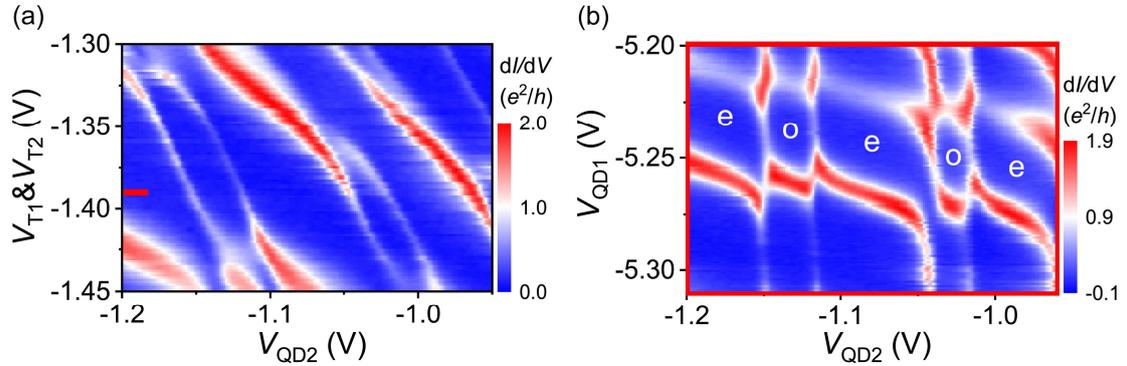

**FIG. S2. Identification of an EOPC in QD2.** (a) $dI/dV$ as a function of $V_{QD2}$ and $V_{T1}\&V_{T2}$, measured at $V_{QD1} = -5.31$ V, $V_{bias} = 0$ mV and $B = 80$ mT. (b) Charge stability diagram of the coupled QD1–QD2 system recorded at the values of $V_{T1}\&V_{T2}$ indicated by the horizontal red bar in (a). The labels "e" and "o" denote the even and odd parity states of QD2, respectively, demonstrating the even–odd parity alternation of QD2 as a function of $V_{QD2}$.

### C. Dependence of supercurrent avoided crossings on $V_{QD1}$

To systematically investigate the relationship between the avoided crossings in the supercurrent and $V_{QD1}$, we measured the zero-bias conductance spectra (Figs. S3(a)-(c)) over different ranges of $V_{QD1}$ while keeping all other parameters constant. We find that, as $V_{QD1}$ becomes more negative, the avoided crossings are progressively suppressed. This behavior is consistent with a reduction in the effective inter-dot coupling, which arises from the diminished coupling between QD1 and the SCs.

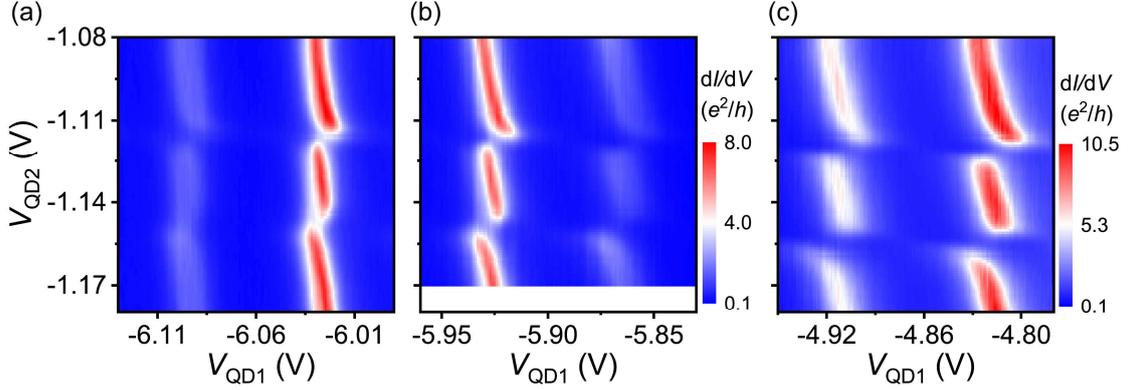

**FIG. S3. Avoided crossings of the supercurrent for different ranges of $V_{QD1}$.** (a), (b), (c) Zero-bias conductance (ZBC) as a function of $V_{QD1}$ and $V_{QD2}$, measured with $V_{T1} = V_{T2} = -1.39$ V.

**D. Charge-number evolution in QD1 and QD2 within the theoretical model of Fig. 2(b)**

Here, we present the calculated charge occupation of QD1 and QD2 as a function of the QDs' chemical potentials ($\mu_{QD1}$ and $\mu_{QD2}$) within the theoretical model corresponding to Fig. 2(b). For details of the theoretical model, please refer to Section III of the Supplemental Material. For QD1, the charge number increments in steps of two as a function of $\mu_{QD1}$ (Fig. S4(a)), and the supercurrent peaks occur at positions where the charge number switches (Fig. 2(b)). In contrast, for QD2, the charge number increments in single-electron steps as a function of $\mu_{QD2}$ (Fig. S4(b)). Crucially, the avoided crossings of the supercurrent coincide with these charge-transition points (Fig. 2(b)).

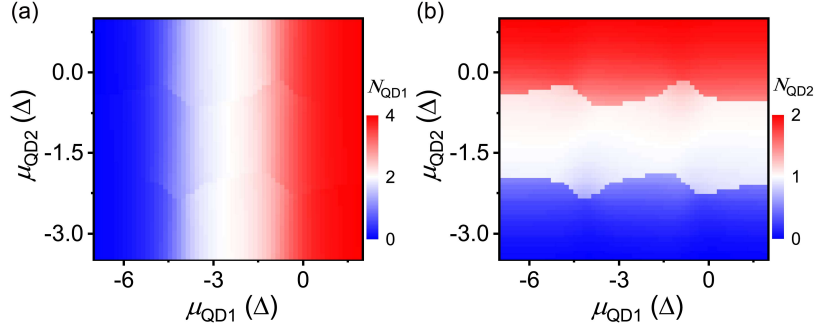

**FIG. S4. Charge-number evolution in QD1 and QD2 within the theoretical model of Fig. 2(b).** (a) Calculated charge number of QD1 as a function of the chemical potentials $\mu_{QD1}$ and $\mu_{QD2}$. (b) Calculated charge number of QD2 as a function of the chemical potentials $\mu_{QD1}$ and $\mu_{QD2}$.

### E. Realization of an EOPC in QD1 at zero magnetic field

The second regime of the joint PC discussed in the main text corresponds to both QD1 and QD2 residing in an EOPC. Here, we demonstrate the realization of an EOPC in QD1 by adjusting the gate voltage $V_{QD1}$ to more negative values. As established in Fig. 1(f), reducing the coupling strength between QD1 and the SCs via decreasing $V_{QD1}$ can drive a PC transition of QD1 from an EPC to an EOPC. Compared with the data shown in Fig. 2(a), we measured the zero-bias conductance spectra at more negative values of $V_{QD1}$ (Fig. S5(a)). As shown in Fig. S5(b), within this $V_{QD1}$ range, QD1 exhibits alternating even-odd parity, confirming that QD1 resides in an EOPC.

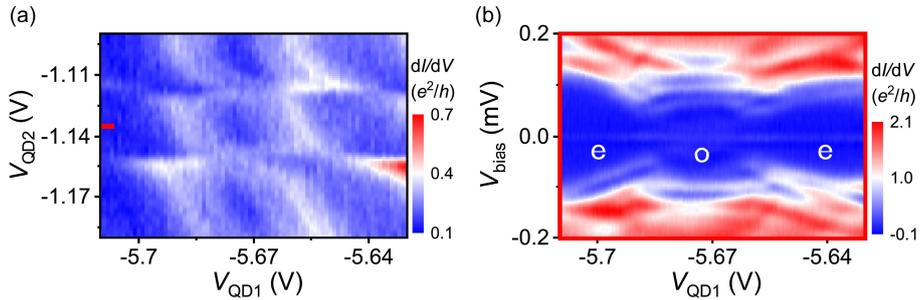

**FIG. S5. QD1 exhibits an EOPC at zero magnetic field.** (a) ZBC as a function of $V_{QD1}$ and $V_{QD2}$ at $B = 0$ mT. The ranges of the gate voltages are identical to those in Fig. 3(b). (b) Spectra of QD1 as a function of $V_{QD1}$ and $V_{bias}$, recorded at a fixed value of $V_{QD2}$ indicated by the horizontal bar in (a). The alternating even–odd parity pattern confirms that QD1 resides in an EOPC.

## F. Evolution of the parity of QD1 with $V_{QD1}$ and $V_{QD2}$

Analogous to Fig. 3(a), we measured the response of the QD1 spectrum to sweeping $V_{QD1}$ or $V_{QD2}$, as shown in Fig. S6(a) (which reproduces the same dataset as Fig. 3(b)). In the off-resonance case (Fig. S6(b)), the odd-parity region is wide. As QD2 approaches resonance (Fig. S6(c)), this region shrinks, indicating the onset of a PC transition. Due to a sub-optimal setting of the fixed $V_{QD2}$ in Fig. S6(c), the complete closure of the odd-parity region is not observed in this specific horizontal cut. However, in Fig. S6(e), as QD2 is tuned into resonance (vertical sweep), the odd-parity regions pinch off completely, as marked by the white arrows. These results demonstrate that the PC transition of QD1 is indeed achieved in the regime of Fig. S6(a) (consistent with the behavior observed in the magnetic-field-induced EOPC regime of Fig. 3(a)).

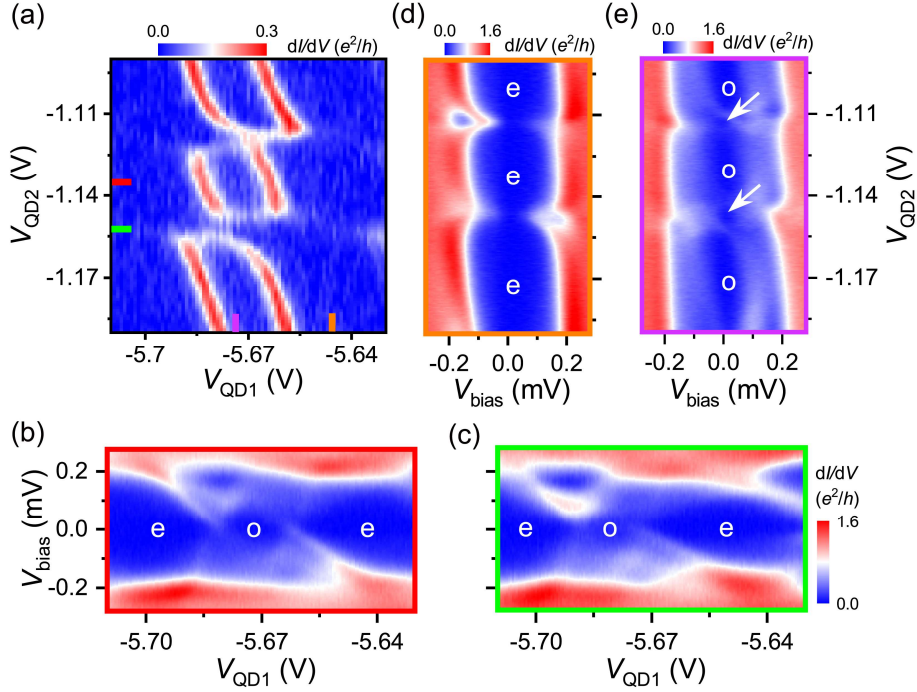

**FIG. S6. Evolution of the parity of QD1 with $V_{QD1}$ and $V_{QD2}$.** (a) The same dataset as Fig. 3(b). (b), (c) d$I$/d$V$ as a function of $V_{bias}$ and $V_{QD1}$, recorded at different fixed values of $V_{QD2}$ indicated by the horizontal red and green bars in (a). (d), (e) d$I$/d$V$ as a function of $V_{bias}$ and $V_{QD2}$, recorded at different fixed values of $V_{QD1}$ indicated by the vertical orange and purple bars in (a). The labels "e" and "o" denote the even and odd parity states of QD1, respectively.

## G. Theoretical analysis of non-local parity control: The role of joint PCs and dominant coupling mechanisms

The main text established the dependence of PC transitions on different joint-PC regimes within the experimentally accessible ECT-dominant limit. Here, we theoretically extend this framework to investigate the symmetry-imposed selection rules by contrasting the two dominant inter-dot coupling mechanisms: particle-conserving ECT and particle-non-conserving CAR.

Extensive parameter-space sampling was performed, and the results are summarized in Table SI. The phase diagrams presented here are representative examples selected from the full dataset. Apart from the theoretical model shown in Fig. S7(m)-(o), which corresponds to Fig. 3(c) in the main text, all other phase diagrams are consistent with the model used in Fig. 4(e). For detailed model descriptions, please refer to Section III of the Supplemental Material. In our simulations, the joint-PC regime is achieved by adjusting the coupling strength between each QD and the SCs ($t_1$ and $t_4$), fixed $t_2 = t_3 = 0.3$. Moreover, the PC of a specific QD is determined by examining its behavior in the phase diagram when another QD is tuned off-resonance. For example, in the phase diagram presented in Fig. S7(a), each QD remains an even-parity state when the other is off-resonance, leading to the identification of a joint EPC-EPC regime. The dominant inter-dot coupling mechanism is tuned by adjusting the chemical potential of the intermediate hybrid segment ($\varepsilon_3$), where $\varepsilon_3 = 0$ corresponds to the CAR-dominant limit and $\varepsilon_3 = -1.5\,\Delta$ ($\Delta$ is set as the energy unit) corresponds to the ECT-dominant limit.

We now analyze specific conditions:

- **Condition I (EPC, EPC; ECT-dominant):** As shown in Figs. S7(a)-(c), both QDs reside in an EPC, with $\varepsilon_3 = -1.5\,\Delta$ (ECT-dominant regime). To clearly examine the PC transition of QD1, we analyze the variation of its particle number ($N_{\text{QD1}}$) with respect to $\mu_{\text{QD1}}$ and $\mu_{\text{QD2}}$ (Fig. S7(b)). Clearly, the PC of QD1 remains unchanged as $\mu_{\text{QD2}}$ is tuned. This confirms the robustness of the EPC under ECT.

- **Condition II (EPC, EPC; CAR-dominant):** Here, both QDs are again in an EPC, but CAR dominates ($\varepsilon_3 = 0$), as shown in Figs. S7(d)-(f). In stark contrast to Condition I, the PC of QD1 can transition from EPC to EOPC (Fig. S7(e)), while the PC of QD2 remains unchanged (Fig. S7(f)).

- **Condition III (EPC, EOPC; ECT-dominant):** This corresponds to Regime I discussed in the main text (Fig. 2). We provide the supplementary theoretical phase diagrams in Figs. S7(g)-(i). Although Fig. S7(g) exhibits a double-arc pattern resembling Fig. 4(e), closer inspection reveals that the PC transition is localized to QD2, while QD1 retains its parity. Therefore, no PC transition is induced in QD1, consistent with the experimental observations in the first regime (Fig. 2).

- **Condition IV (EPC, EOPC; CAR-dominant):** As shown in Figs. S7(j)-(l), no distinct PC transition is observed for QD1 (Fig. S7(k)).

- **Condition VI (EOPC, EOPC; CAR-dominant):** Figs. S7(m)-(o) correspond to the counterpart of Regime II (Fig. 3) but in the CAR limit. Compared with Fig. 3(c), while the PC transition of QD1 persists, the orientation of the avoided crossing rotates (from anti-diagonal to diagonal).

- **Condition VIII (EOPC, EPC; CAR-dominant):** The coupling parameters $t_1$ and $t_4$ for the two QDs in Fig. S7(p) are exactly the opposite of those in Fig. S7(j) (Condition IV). Under this CAR-dominant condition, while distortions appear in the phase diagram, no PC transition is observed for QD1. This contrasts with the ECT case (main text Regime III; Fig. 4), where a PC transition occurs.

- **Conditions V, and VII:** These two conditions correspond to the main text Regimes II and III, respectively. They have been discussed in detail in the main text, and are therefore not repeated here.

We briefly clarify the identification of the ECT-dominant mechanism throughout the experiments. For Regime II (Condition V), the dominance of ECT is unambiguously identified by the anti-diagonal orientation of the avoided crossings, as the CAR-dominant limit would yield a diagonal pattern. In Regime III (Condition VII), the experimental observation of a PC transition (Fig. 4) is consistent with the theoretical simulation in the ECT-dominant limit (Fig. 4(e)). This stands in contrast to the CAR-dominant case (Condition VIII), where the selection rules prohibit such a transition. Finally, regarding Regime I (Condition III), since the chemical potential of the intermediate hybrid segment—which governs the competition between ECT and CAR—remains largely unperturbed across the explored gate-voltage space, Regime I is expected to operate under the same ECT-dominant mechanism established for Regimes II and III.

In summary, Table I in the main text establishes the selection rules in the ECT-dominant limit. Table SI extends this framework to the CAR-dominant limit. Together, they provide a comprehensive phase diagram demonstrating that non-local parity control is governed by universal constraints imposed by the joint PC and the symmetry of the inter-dot coupling.

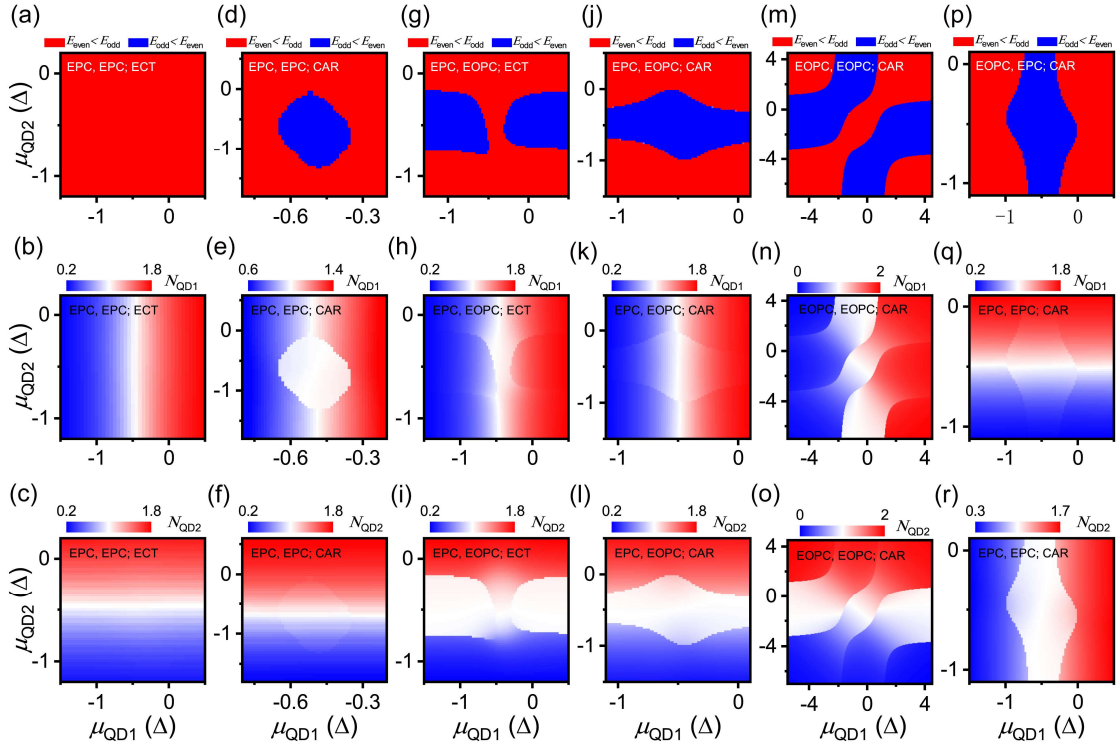

**FIG. S7. Phase diagrams obtained from theoretical simulations under different parameter sets.** Except for panel (m)-(o), which corresponds to the theoretical model used in Fig. 3(c), all other phase diagrams are based on the model used in Fig. 4(e). Some parameters in Figs. (a)-(l), (p)-(r) are identical: $\varepsilon_1 = \varepsilon_2 = \varepsilon_4 = \varepsilon_5 = 0$; $U_1 = U_2 = 1$; $E_z = 0$; $t_2 = t_3 = 0.3$; $t_{sc}/t_{sf} = 5$ & $t_{sc}^2 + t_{sf}^2 = 1$. (a)-(c) Calculated with parameters: $\varepsilon_3 = -1.5$; $t_1 = 0.6$; $t_4 = 0.6$. (d)-(f) Calculated with parameters: $\varepsilon_3 = 0$; $t_1 = 0.5$; $t_4 = 0.62$. (g)-(i) Calculated with parameters: $\varepsilon_3 = -1.5$; $t_1 = 0.56$; $t_4 = 0.48$. (j)-(l) Calculated with parameters: $\varepsilon_3 = 0$; $t_1 = 0.56$; $t_4 = 0.48$. (m)-(o) Calculated with parameters: $\varepsilon_1 = \varepsilon_2 = 0$; $U_1 = 1$; $U_2 = 2.5$; $E_z = 1$; $t = \sqrt{2}$; $\delta = 1.7$; $t_{sc}/t_{sf} = 1.25$ & $t_{sc}^2 + t_{sf}^2 = 1$; $\delta_{sc}/\delta_{sf} = 1.25$ & $\delta_{sc}^2 + \delta_{sf}^2 = 1$. (p)-(r) Calculated with parameters: $\varepsilon_3 = 0$; $t_1 = 0.48$; $t_4 = 0.56$.

**Table SI. Summary of non-local control of the PC of QD1 under different joint-PC regimes and dominant inter-dot coupling mechanisms.**

| Conditions | PC of QD1 | PC of QD2 | Dominant inter-dot coupling mechanism | PC transition of QD1 | Figures |
|---|---|---|---|---|---|
| Condition I | EPC | EPC | ECT dominant | No | Figs. S7(a)-(c) |
| Condition II | | | CAR dominant | Yes | Figs. S7(d)-(f) |
| Condition III | EPC | EOPC | ECT dominant | No | **Fig. 2** and Figs. S7(g)-(i) |
| Condition IV | | | CAR dominant | No | Figs. S7(j)-(l) |
| Condition V | EOPC | EOPC | ECT dominant | Yes | **Fig. 3** |
| Condition VI | | | CAR dominant | Yes | Fig. S7(m)-(o) |
| Condition VII | EOPC | EPC | ECT dominant | Yes | **Fig. 4** |
| Condition VIII | | | CAR dominant | No | Fig. S7(p)-(r) |

Note: PC, parity configuration; QD, quantum dot; EPC, even parity configuration; EOPC, even–odd parity configuration; ECT, elastic co-tunneling; CAR, crossed Andreev reflection.

## III. Theoretical model

### A. Transport model corresponding to Fig. 2(b)

We describe the simulation of the supercurrent of QD1 shown in Fig. 2(b). The SC-QD-SC-QD-SC Hamiltonian is given by

$$H = \sum_{j=1,3,5} H_S^j + \sum_{j=2,4} H_{QD}^j + \sum_{j=1}^4 H_T^j, \quad (1)$$

For simplicity, we assume a single-level SC:

$$H_S^j = \sum_{\sigma=\uparrow,\downarrow} \varepsilon_j n_{j,\sigma} + (\Delta e^{i\varphi_j} c_{j,\uparrow}^\dagger c_{j,\downarrow}^\dagger + h.c.), \quad (2)$$

$H_S^j$ is the Hamiltonian of the SC with the odd index $j$, $\varepsilon_j$ is the normal-state energy, $\Delta$ denotes the induced superconducting pairing potential, and $\varphi_j$ represents the phase. Here, $n_{j,\sigma} = c_{j,\sigma}^\dagger c_{j,\bar\sigma}$, $c_{j,\sigma}^\dagger (c_{j,\bar\sigma})$ is the electron creation (annihilation) operator, and $\bar\sigma$ means the spin opposite to $\sigma$.

Next, we consider the orbits in two QDs. QD1 is modeled with two energy levels, whereas QD2 is modeled with a single level, reflecting the necessity to include up to four-electron occupation in QD1.

$$H_{QD}^2 = \sum_{k=1,2} \sum_{\sigma=\uparrow,\downarrow} (\varepsilon_{2,k} + \eta_\sigma E_Z) n_{2,k,\sigma} + \sum_{k=1,2} U_{2,k} n_{2,k,\uparrow} n_{2,k,\downarrow} + U_c (n_{2,1,\uparrow} +$$

$$n_{2,1,\downarrow})(n_{2,2,\uparrow} + n_{2,2,\downarrow}), \tag{3}$$

$$H_{QD}^4 = \sum_{\sigma=\uparrow,\downarrow}(\varepsilon_{4,1} + \eta_\sigma E_Z)n_{4,1,\sigma} + U_{4,1}n_{4,1,\uparrow}n_{4,1,\downarrow}, \tag{4}$$

$H_{QD}^2$ is the Hamiltonian of QD1 and $H_{QD}^4$ is the Hamiltonian of QD2, $\varepsilon_{j,k}$ is the on-site energy of level $k$, $\eta_\uparrow = +1$ ($\eta_\downarrow = -1$), $E_Z$ denotes the Zeeman spin splitting, $U_{j,k}$ represents the Coulomb repulsion, and $U_c$ represents the Coulomb repulsion between the two levels of QD1. Here, $n_{j,k,\sigma} = c_{j,k,\sigma}^\dagger c_{j,k,\bar\sigma}$, $c_{j,k,\sigma}^\dagger(c_{j,k,\bar\sigma})$ is the electron creation (annihilation) operator.

At last, we consider coupling SCs to different levels of the adjacent QDs respectively.

$$H_T^1 = t_1 \sum_{k=1,2}\sum_{\sigma=\uparrow,\downarrow}(t_{sc}c_{1,\sigma}^\dagger c_{2,k,\sigma} + \eta_\sigma t_{sf}c_{1,\sigma}^\dagger c_{2,k,\bar\sigma} + h.c.). \tag{5.1}$$

$$H_T^2 = t_2 \sum_{k=1,2}\sum_{\sigma=\uparrow,\downarrow}(t_{sc}c_{2,k,\sigma}^\dagger c_{3,\sigma} + \eta_\sigma t_{sf}c_{2,k,\sigma}^\dagger c_{3,\bar\sigma} + h.c.). \tag{5.2}$$

$$H_T^3 = t_3 \sum_{k=1}\sum_{\sigma=\uparrow,\downarrow}(t_{sc}c_{3,\sigma}^\dagger c_{4,k,\sigma} + \eta_\sigma t_{sf}c_{3,\sigma}^\dagger c_{4,k,\bar\sigma} + h.c.). \tag{5.3}$$

$$H_T^4 = t_4 \sum_{k=1}\sum_{\sigma=\uparrow,\downarrow}(t_{sc}c_{4,k,\sigma}^\dagger c_{5,\sigma} + \eta_\sigma t_{sf}c_{4,k,\sigma}^\dagger c_{5,\bar\sigma} + h.c.). \tag{5.4}$$

$H_T^j$ is the Hamiltonian of nearest-neighbor coupling, where $t_j$ denotes the coupling strength, and $t_{sc}$ ($t_{sf}$) represents the spin-conserving (spin-flipping) component.

The current operator flowing through the left SC site is defined as

$$i\hbar j_1 = [n_1, H_S^1] = [n_1, H_0]$$

where $j_1$ is proportional to the supercurrent.

Figure 2(b) is calculated using the parameters: $\varepsilon_1 = \varepsilon_2 = \varepsilon_3 = \varepsilon_4 = \varepsilon_5 = 0$; $U_{2,1} = U_{2,2} = 1$; $U_{4,1} = 2.5$; $U_c = 1.5$; $t_j = 0.5$; $t_{sc}/t_{sf} = 3$ & $t_{sc}^2 + t_{sf}^2 = 1$; $E_Z = 0$. Throughout this work, we choose $\Delta$ as the energy unit.

**B. Model of the double-QD phase diagram corresponding to Fig. 3(c)**

We next explain the theoretical model used to calculate the phase diagram shown in Fig. 3(c). To obtain a more intuitive physical picture (for instance, using $t > \delta$ to indicate that the system is ECT dominant, etc.), we simplified the Hamiltonian and equivalent the influence of the SC to the ECT process and CAR process with different spin configurations between two QDs. In this case, the double-QD Hamiltonian is given by

$$H_{DQD} = \sum_{j=2,4} H_{QD}^j + H_C. \tag{6}$$

It is worth noting that in Fig. 3(c), based on the experimental data, we assume that QD1 has only one level.

$$H_{QD}^j = \sum_{\sigma=\uparrow,\downarrow}(\varepsilon_j + \eta_\sigma E_Z)n_{j,\sigma} + U_j n_{j,\uparrow}n_{j,\downarrow}, \quad (7)$$

$$H_C = t\sum_{\sigma=\uparrow,\downarrow}(t_{sc}c_{1,\sigma}^\dagger c_{2,\sigma} + \eta_\sigma t_{sf}c_{1,\sigma}^\dagger c_{2,\bar{\sigma}} + h.c.) + \delta\sum_{\sigma=\uparrow,\downarrow}(\delta_{sc}c_{1,\sigma}^\dagger c_{2,\bar{\sigma}}^\dagger + \eta_\sigma \delta_{sf}c_{1,\sigma}^\dagger c_{2,\bar{\sigma}}^\dagger + h.c.). \quad (8)$$

Here, $n_{j,\sigma} = c_{j,\sigma}^\dagger c_{j,\bar{\sigma}}$, $c_{j,\sigma}^\dagger(c_{j,\bar{\sigma}})$ is the electron creation (annihilation) operator, and $\bar{\sigma}$ means the spin opposite to $\sigma$. $H_{QD}^2$ is the Hamiltonian of QD1 and $H_{QD}^4$ is the Hamiltonian of QD2, $\varepsilon_j$ is the on-site energy, $E_Z$ denotes the Zeeman spin splitting and $U_j$ represents the Coulomb repulsion. $H_C$ is the effective coupling, $t$ denotes the effective coupling strength of elastic co-tunneling, $t_{sc}$ ($t_{sf}$) represents the spin-conserving (spin-flipping) component; $\delta$ denotes the effective coupling strength of crossed Andreev reflection, and $\delta_{sc}$ ($\delta_{sf}$) represents the spin-conserving (spin-flipping) component.

Figure 3(c) is calculated with parameters: $\varepsilon_1 = \varepsilon_2 = 0$; $U_1 = 1$; $U_2 = 2.5$; $E_z = 1$; $t = \sqrt{2}$; $\delta = 1.05$; $t_{sc}/t_{sf} = 1.25$ & $t_{sc}^2 + t_{sf}^2 = 1$; $\delta_{sc}/\delta_{sf} = 1.25$ & $\delta_{sc}^2 + \delta_{sf}^2 = 1$. Again, we choose $\Delta$ as the energy unit.

### C. Model of the phase diagram corresponding to Fig. 4(e)

We now show the model used to obtain the phase diagram shown in Fig. 4(e). Compared to Fig. 3(c), to better correspond with the experimental system, we consider the influence of the SCs in more detail, building upon the double QDs setup. The SC-QD-SC-QD-SC Hamiltonian is

$$H = \sum_{j=1,3,5} H_S^j + \sum_{j=2,4} H_{QD}^j + \sum_{j=1}^4 H_T^j, \quad (9)$$

where

$$H_S^j = \sum_{\sigma=\uparrow,\downarrow}\varepsilon_j n_{j,\sigma} + (\Delta e^{i\varphi_j}c_{j,\uparrow}^\dagger c_{j,\downarrow}^\dagger + h.c.), \quad (10)$$

$$H_{QD}^j = \sum_{\sigma=\uparrow,\downarrow}(\varepsilon_j + \eta_\sigma E_Z)n_{j,\sigma} + U_j n_{j,\uparrow}n_{j,\downarrow}, \quad (11)$$

$$H_T^j = t_j \sum_{\sigma=\uparrow,\downarrow}(t_{sc}c_{j,\sigma}^\dagger c_{j+1,\sigma} + \eta_\sigma t_{sf}c_{j,\sigma}^\dagger c_{j+1,\bar{\sigma}} + h.c.). \quad (12)$$

Here, $n_{j,\sigma} = c_{j,\sigma}^\dagger c_{j,\bar{\sigma}}$, $c_{j,\sigma}^\dagger(c_{j,\bar{\sigma}})$ is the electron creation (annihilation) operator and $\bar{\sigma}$ means the spin opposite to $\sigma$. $H_S^j$ is the Hamiltonian of the SC with the odd index $j$, $\Delta$ denotes the induced superconducting pairing potential, and $\varphi_j$ represents the phase. $H_{QD}^2$ is the Hamiltonian of QD1 and $H_{QD}^4$ is the Hamiltonian of QD2, $\varepsilon_j$ is the on-site energy, $\eta_\uparrow = +1\,(\eta_\downarrow = -1)$, $E_Z$ denotes the Zeeman spin splitting and $U_j$ represents the Coulomb repulsion. $H_T^j$ is the nearest-neighbor coupling, where $t_j$ denotes the coupling strength, and $t_{sc}$ ($t_{sf}$) represents the spin-conserving (spin-flipping) component. Compared with the model used for Fig. 2(b), the present model includes only a single level in each QD, while retaining the same form of nearest-neighbor coupling.

Figure 4(e) is calculated using the parameters: $\varepsilon_1 = \varepsilon_2 = \varepsilon_4 = \varepsilon_5 = 0$; $\varepsilon_3 = -1.5$; $U_1 = U_2 = 1$; $E_z = 0$; $t_2 = t_3 = 0.3$; $t_1 = 0.48$; $t_4 = 0.56$; $t_{sc}/t_{sf} = 5$ & $t_{sc}^2 + t_{sf}^2 = 1$. We choose $\Delta$ as the energy unit.